\documentclass[showpacs,preprintnumbers,english,twocolumn,amsmath,amssymb
]{revtex4-1}
\usepackage[T1]{fontenc}
\usepackage[latin1]{inputenc}
\usepackage{graphicx}
\usepackage{amssymb}
\usepackage{verbatim}
\usepackage{epic,eepic}
\usepackage{babel}
\DeclareMathAlphabet{\mathpzc}{OT1}{pzc}{m}{it} 

\begin{document}

\title{Constraining the Symmetry Term in the Nuclear Equation of State at Sub-Saturation Densities and Finite Temperatures}

\author{P.~Marini$^{1,}$}\email{pmarini@comp.tamu.edu}
\author{A.~Bonasera$^{1,2}$}
\author{A.~McIntosh$^{1}$}
\author{R.~Tripathi$^{1,3}$}
\author{S.~ Galanopoulos$^{1,}$}
   \altaffiliation{Present address: Greek Army Academy, Department of Physical Science, Athens, Greece.}
\author{K.~Hagel$^{1}$}
\author{L.~ Heilborn$^{1,4}$}
\author{Z.~Kohley$^{1,4,}$}
    \altaffiliation{Present address: National Superconducting Cyclotron Laboratory, Michigan State University, East Lansing, Michigan 48824, USA.}
\author{L.~W.~May$^{1,4}$}
\author{M.~ Mehlman$^{1,5}$}
\author{S.N. Soisson $^{1,2}$}
\author{G.~A.~Souliotis$^{1,6}$}
\author{D.V. Shetty$^{1}$}
   \altaffiliation{Present address: Physics Department, Western Michigan University, Kalamazoo, MI 49008, USA.}
\author{W.~ B.~ Smith$^{1}$}
\author{B.C. Stein$^{1,2}$}
\author{S.~Wuenschel$^{1,4}$}
\author{S.~J.~Yennello$^{1,4}$}

\affiliation{
$^{1}$Cyclotron Institute, Texas A\&M University, College Station, TX-77843, USA\\
$^{2}$Laboratori Nazionali del Sud, INFN, via Santa Sofia, 62, 95123 Catania, Italy\\
$^{3}$Radiochemistry Division, Bhabha Atomic Research Center, Mumbai, India\\
$^{4}$Chemistry Department, Texas A\&M University, College Station, TX-77843, USA\\
$^{5}$Physics and Astronomy Department, Texas A\&M University, College Station, Texas 77843, USA\\
$^{6}$Laboratory of Physical Chemistry, Department of Chemistry, National and Kapodistrian University of Athens, 15771 Athens, Greece}

\date{\today}

\begin{abstract}
Methods of extraction of the symmetry energy (or enthalpy) coefficient to temperature ratio from isobaric and isotopic yields of fragments produced in Fermi-energy heavy-ion collisions are discussed.  We show that the methods are consistent when the hot fragmenting source is well characterized and its excitation energy and isotopic composition   are properly taken into account.  The results are independent of the mass number of the detected fragments, which suggests that their fate is decided very early in the reaction.
\end{abstract}


\maketitle

\section{Introduction}
In the past few years the 
importance of the symmetry energy term in the nuclear equation of state has stimulated a growing interest in isospin effects in nuclear reactions. Understanding the properties of asymmetric nuclear matter both at normal densities and at densities away from the saturation density has an important impact on the study of the nuclear structure close to the drip lines \cite{brown2000} and on the study of astrophysical processes \cite{lattimer01}.

Recent measurements of the giant dipole \cite{trippa08}, Pygmy dipole \cite{klimkiewicz07} and giant monopole \cite{li07} resonances in neutron-rich nuclei, neutron/proton emission \cite{famiano06}, isospin diffusion \cite{tsang2004} and fragment isotopic ratio \cite{tsang2001,iglio06} have provided initial constraints on the density dependence of the symmetry energy at sub-saturation densities. Refinement of these measurements with both stable and rare isotope beams in the near future will provide further stringent constraints. New crucial experimental constraints on the symmetry energy at supra-saturation densities are expected from the recent measurement on neutron-proton elliptic flow performed at GSI \cite{proposal_paolo}. Nevertheless the conclusions are model dependent, different analysis methods provide different results and a consistent description of all available experimental data employing one type of equation of state is still lacking.

In this work, among the experimental observables commonly used to explore the symmetry energy at sub-saturation densities, we will concentrate on yield ratios of fragment produced in multifragmentation processes at Fermi energies.
Different methods to extract the ``symmetry energy'' coefficient 
from this observable have been proposed. Among those, we will focus on three seemingly different approaches:  isoscaling 
\cite{xu2000,tsang2001_2,botvina2002,ono03,souliotis04,lefevre05},  m-scaling \cite{huang2010Mscaling} and  isobaric yield ratio method \cite{huang2010IYR}, and we will show that consistent experimental results can be extracted provided that the properties of the hot fragmenting source, such as its isospin composition and excitation energy, are properly taken into account. The paper is structured as follow: section \ref{sec:theoretical_background}  briefly describes the three methods. Sections \ref{sec:motivation} and \ref{sec:experiment}  present the motivation that led to this work and the experimental apparatus, respectively. The importance of the source reconstruction and the results are presented in sections \ref{sec:QP rec} and \ref{sec: comparison} respectively, while section \ref{sec:conclusions} concludes.

\section{Theoretical background}\label{sec:theoretical_background}
The study of the multifragmentation process in violent heavy-ion collisions  at Fermi energies is important for the investigation of the symmetry energy. During the multifragmentation process, which has been related to a nuclear phase transition \cite{bonasera94, bonasera2000, borderie2008,trautmann2005}, subsaturation  density may be achieved  \cite{kowalski07}.
 Free energies play an important role in mixed phase environments. Indeed the isotopic distribution of fragments produced in such collisions is governed by the free energy at the pressure and temperature of the fragmenting source. Thus, assuming that the fragment production is governed by a purely statistical process at constant temperature T and pressure $P$, the yield of a fragment, with N neutrons and Z protons,  $Y(N,Z,T,P)$, can be related to the nuclear Gibbs free energy $G(N,Z,T,P)$ \cite{ono03}:
\begin{equation}\label{yield}
\begin{array}{c}
Y(N,Z,T,P)=\\
\vspace*{-2mm}\\
 Y_{0} A^{-\tau} exp \left\lbrace -\frac{G(N,Z,T,P)}{T}+\frac{\mu_{n}}{T}N+\frac{\mu_{p}}{T}Z\right\rbrace
\end{array}
\end{equation}
where $Y_{0}$ is a constant, $\mu_{n}$ and $\mu_{p}$ are the neutron and proton chemical potentials, $T$ is the temperature  of the emitting source. The factor $A^{-\tau}$ originates from the entropy of the fragment, i.e. the  Fisher entropy \cite{minich82,belkacem95}.
The use of the Gibbs free energy \cite{ono2004} depends on the validity of the assumption of fragment production via an equilibrium mechanism at constant pressure. Other assumptions, and therefore other thermodynamic state functions, could be used \cite{das2005,bondorf95}, but it is undoubtedly true that the true fragment production scheme requires a kinetic treatment. For instance if the volume is kept constant (freeze-out hypothesis), then the Helmholtz free energy, $F(N,Z,T,V)$, should be used. This ambiguity casts some doubts on the derived quantity, i.e. symmetry energy or enthalpy.

Within a liquid drop description, the nuclear free energy, $F(N,Z,T,V)$, can be parametrized as a sum of the bulk, surface, Coulomb and symmetry free energy contributions. 
The symmetry  energy term  
is usually expressed as 
\cite{chen10, tsang2001_2}
\begin{equation}\label{Gsym}
E_{sym}(N,Z,T,V)=C_{sym}(T,V)\frac{(N-Z)^{2}}{A},
\end{equation}
where $C_{sym}(T,V)$ is the symmetry energy coefficient, $V$ is the volume and $A=N+Z$.
Similarly, starting from the Gibbs free energy, $G(N,Z,T,P)$, the symmetry enthalpy term could be espressed introducing a symmetry enthalpy coefficient $C_{hsym}$, depending on $(T,P)$.
The logic employed in this work, which assumes an equilibrium process at constant pressure \cite{ono03}, will lead to the extraction of the symmetry enthalpy. The assumption of a constant volume \cite{tsang2001_2} would have lead to the extraction of the symmetry energy.
Experimentally whether the equilibrium process takes place at constant pressure or volume (freeze-out hypothesis) is not determined, therefore the ambiguity on the extracted quantity is retained.
However some estimates show that the difference between the two quantities should be small below the critical point of the phase transition \cite{sobotka2011}. Above it, the $PV$ product becomes finite, and the two quantities might significantly differ. Nevertheless in Ref. \cite{das2005} it is shown that the canonical and grand-canonical ensembles predict similar results if one is only interested in the ratio of population of two adjacent isotopes, on which we will focus in this work.
Keeping in mind this ambiguity, from now on we will refer to the experimentally extracted quantity as the ``symmetry energy".\\

We now give a brief review of the three methods that we will use in our experimental analysis.\\
It has often been experimentally observed that the ratio of the  yields of a fragment with $N$ neutrons and $Z$ protons produced in two similar reaction systems with different neutron-to-proton ratios  is exponential in $N$ and $Z$ \cite{xu2000,tsang2001_2,botvina2002,ono03,souliotis04,lefevre05}.
In the grand canonical approximation, using Eq. \ref{yield} this ratio  can be written as
\begin{equation}\label{R12isoscaling}
\left.
\begin{array}{c}
R_{21}(N,Z,T,P)=\frac{Y_{2}(N,Z,T,P)}{Y_{1}(N,Z,T,P)}\\
\vspace*{-2mm}\\
= \frac{Y_{0,2}}{Y_{0,1}} exp \left\lbrace \left[   \left(\mu_{n,2}-\mu_{n,1}\right)N+\left(\mu_{p,2}-\mu_{p,1}\right)Z \right]  /T \right \rbrace  \\
\vspace*{-2mm}\\
= C exp (\alpha N+\beta Z)
    \end{array}\right.   
\end{equation}
assuming that the thermodynamic state points 
of the two equilibrated sources in the two reactions are the same. (The indices $1$ and $2$ denote the neutron-poor and neutron-rich system, respectively).
This relation, known as  isoscaling, has been found to describe the measured ratios over a wide range of complex fragments and light particles rather well \cite{tsang2001_2} and to be a phenomenon common to many different types of heavy-ion reactions \cite{botvina2002,souliotis04,souliotis2006,galanopoulos2010,lefevre05,tsang2001}.
 This suggests that free energy components sensitive to the neutron-proton concentration differences play a key role in the fragment formation process.\\

Recently, the Modified Fisher Model of Ref. \cite{minich82} has been used to interpret  multifragmentation data and extract information on the symmetry energy \cite{huang2010Theory, huang2010Mscaling, huang2010IYR}. In this model, the  free energy per particle is modified as \cite{huang2010Theory}:
\begin{equation}\label{FTvsG}
\Psi(m_{f},A,T,H) \leftrightarrow \frac{G(N,Z,T,P)-\mu_{n}N-\mu_{p}Z}{A}
\end{equation}
where $m_{f}$ ($=\frac{N-Z}{A}$) is the relative isospin asymmetry of the fragment. This is the Landau free energy. The ansatz here is that near a critical point all the dependencies of the free energy are contained in the order parameter $m_{f}$ and its conjugate field $H$ \cite{huang_book,huang2010Theory,huang2010Mscaling}. In principle we do not need to specify if the system is at constant volume or pressure, as before.
Comparing to Eq. \ref{yield}, we should notice that  $\Psi(m_{f},A,T,H)$ includes the neutron and proton chemical potentials. Within this approach, the ratio of the  free energy per particle to the temperature near the critical point is described by the expansion \cite{huang_book,bonasera2008,huang2010Theory}:
\begin{equation}\label{FT}
\left.
\begin{array}{c}
\frac{\Psi(m_{f},A,T,H)}{T}=\\
\vspace*{-2mm}\\
\frac{1}{2}am_{f}^{2}+\frac{1}{4}bm_{f}^{4}+\frac{1}{6}cm_{f}^{6}-\frac{H}{T}m_{f}+\mathpzc{O} (m_{f}^{8}).
\end{array}\right.   
\end{equation}
The parameters $a$, $b$ and $c$, which depend on $T$ and $\rho$ \cite{huang2010Theory}, are used for fitting. We stress here that the ``external field'' in our case might be attributed to the difference in neutron and proton chemical potentials of the source \cite{huang_book, huang2010Mscaling}. 
Indeed in the limit where $\mu_{p}=-\mu_{n}$, combining Eqs. \ref{FTvsG} and \ref{FT}, we obtain $2H=\mu_{n}-\mu_{p}$.
Other terms, such as the volume, surface, Coulomb and pairing contributions, might be negligible near the phase transition, as suggested by experimental data \cite{huang2010Theory}.

An exponential dependence on $m_{f}$ has been experimentally observed for the ratio, $R_{21}$, of the  yields of a fragment ($m_{f}$, $A$) 
produced in two similar reactions with different neutron-to-proton ratios \cite{huang2010Mscaling}.
This scaling has been referred to as \textit{m-scaling}, to distinguish it from the known isoscaling.
Indeed the fragment yield ratio between two systems  at the same thermodynamic state point 
depends only on the external field $H/T$:
\begin{equation}\label{R12mirror}
R_{21}(m_{f},A,T,H)=C exp(\frac{\Delta H}{T}m_{f}\,A)
\end{equation}
where $\Delta H/T=H_{2}/T-H_{1}/T$. Comparing Eq. \ref{R12isoscaling} and \ref{R12mirror} and assuming that $\alpha=-\beta$, we obtain $\frac{\Delta H}{T}=\alpha$, as has been suggested in Ref. \cite{huang2010Mscaling}. As w will show that this relation is approximately satisfied in our experimental data, ignoring the isospin symmetry breaking Coulomb effects is a reasonable approximation for light nuclei \cite{huang2010Theory}.\\

Following the statistical interpretation of the isoscaling with SMM \cite{botvina2002} and with the expanding emitting source model \cite{tsang2001_2}, it has been proposed that the symmetry energy coefficient can be extracted  from the measured isoscaling parameters through the approximate formula \cite{tsang2001_2}:
\begin{equation} \label{csym}
\frac{C_{sym}(T)}{T}=\frac{\alpha}{4\Delta}.
\end{equation}
The two models of Refs. \cite{tsang2001_2,botvina2002} are based on different assumptions, therefore we restrain from stating explicitly the  $C_{sym}$ dependence on $V$ or $P$, which is model dependent.

In the literature a variety of definitions of the quantity $\Delta$ has been proposed \cite{tsang2001_2,ono03}. Among those we will focus on two commonly used expressions and on one recently suggested.

In the context of SMM and the expanding emitting source model, $\Delta$ is the difference of the asymmetries of the equilibrated emitting sources, defined as \cite{tsang2001_2}:
\begin{equation}\label{eq:deltaSource}
\Delta_{source} =  \left[ \left(\frac{Z}{ A}\right)^{2}_{1}-\left(\frac{Z}{ A }\right)^{2}_{2}\right]
\end{equation}

 In general, due to the phenomenon of fractionation, fragments may not have the same isospin ratio of the fragmenting source and this could affect the value of $C_{sym}$ obtained from the isoscaling coefficient.

A study of the isospin fractionation and the isoscaling with AMD simulations \cite{onoAMD} has shown a linear relation between $\alpha$ and $(Z/A)^{2}$ of fragments. The quantity $\Delta$ has then been expressed as
\begin{equation}\label{eq:delta liquido}
\Delta_{liquid}(Z)= \left[\left(\frac{Z}{\langle A\rangle}\right)^{2}_{1}-\left(\frac{Z}{\langle A\rangle}\right)^{2}_{2}\right]
\end{equation}
where $Z/\langle A\rangle$ is  the proton fraction of the most probable isotope for  fragments of a given $Z$ with $A>4$ (``liquid part'') \cite{ono03}.

More recently, R.~Tripathi et al. \cite{rahul} have shown, in an investigation of the nuclear phase transition using the Landau Free energy approach, that the position of the central minimum of the free energy is related to the average isospin asymmetry of the fragments produced in each event, excluding neutrons and protons, $\overline{m_{f}}$. This suggests $\Delta$ may be written as
\begin{equation}\label{eq:delta mf}
\Delta_{\langle m_{f}\rangle}= \left(\frac{1-\langle \overline{m_{f}}\rangle}{2}\right)^{2}_{1}-\left(\frac{1-\langle \overline{m_{f}}\rangle}{2}\right)^{2}_{2},
\end{equation}
where $\langle \overline{m_{f}}\rangle$ is the event average of the fragment relative isospin asymmetry.
\begin{equation}
\langle \overline{m_{f}}\rangle = 
\dfrac{1}{N} \sum_{i=1}^{N} \left[ \sum_{j=1}^{K} \frac{ m_{f_{j}}}{K_{i}} \right]
\end{equation}
where N is the number of events and K the event multiplicity excluding neutrons and protons.

We would like to stress here that the definitions of $\Delta$ proposed in Eq. \ref{eq:deltaSource}, \ref{eq:delta liquido} and \ref{eq:delta mf}  differ significantly. While Eq. \ref{eq:deltaSource} relates $\Delta$ to the characteristics of the two sources, Eqs. \ref{eq:delta liquido} and \ref{eq:delta mf} express $\Delta$ as a function of the characteristics of the fragments. Moreover Eq. \ref{eq:delta mf}, as opposed to Eq. \ref{eq:delta liquido}, takes into account, event-by-event, the  multiplicity in addition to the composition of the fragments. Since the determination of $C_{sym}/T$ from scaling parameters depend on $\Delta$, the correct determination of $\Delta$ is critical.\\

Recently, a different method to determine $C_{sym}/T$ has been proposed, which does not depend on the definition of $\Delta$.
Within the  Modified Fisher Model \cite{minich82} it has been  shown that the isotope yield ratio  
between two isobars differing by 2 units in the neutron excess $I = N-Z$ 
and produced by the same source can be written as \cite{huang2010IYR}:
\begin{equation}\label{R(I+2,I)}
\left.
\begin{array}{c}
R(I+2,I,A,T,\rho)=\frac{Y(I+2,A,T,\rho)}{Y(I,A,T,\rho)}  \\
\vspace*{-2mm}\\
=exp \left\lbrace \frac{W(I+2,A,T,\rho)-W(I,A,T,\rho)+ (\mu_{n}-\mu_{p})}{T}+ \right. \\
\vspace*{-2mm}\\
\left. +S_{mix}(I+2,A)-S_{mix}(I,A) 
\right\rbrace
     \end{array}\right.   
\end{equation}
where 
$W(I, A, T, \rho)$ is the free energy of the cluster at temperature T, which in Ref. \cite{huang2010IYR} was approximated by the generalized Weisz\"{a}cker-Bethe semi-classical mass formula \cite{weizsacker35,bethe36}, and 
$S_{mix}(I,A)$ is the cluster mixing entropy. 
Apart the change of the notation for the free energy, we stress that a similar formula could be derived at constant pressure, thus the ambiguity discussed above remains also in this case.
Note that the variable $I$, introduced for consistency with Ref. \cite{huang2010IYR}, could be expressed as a function of the isospin asymmetry $m_{f}$ as $I=m_{f}\, A$. The authors show that the symmetry energy coefficient to temperature ratio can be expressed as:
\begin{equation} \label{cymsIYRLandau}
\left.
\begin{array}{c}
\frac{C_{sym}(T)}{T}\approx -\frac{A}{8}\left[ lnR(3,1,A) -\right.\\
\vspace*{-2mm}\\
\left. \hspace{2cm}-lnR(1,-1,A)- \delta(3,1,A)\right]
 \end{array}\right.   
\end{equation}
where $\delta(3,1,A)$ is the difference in the mixing entropies and can be neglected, being rather small compared to the other terms in Eq. \ref{cymsIYRLandau} \cite{huang2010IYR}.

Equation \ref{cymsIYRLandau} can  be derived within a Landau free energy approach, which is used here to show that the dependence of $C_{sym}/T$ on the source characteristics, contained in $H/T$, cancels out to first order. The yield of each fragment can be expressed as a function of its free energy per particle $F$, as in Eq. \ref{FT}.
Neglecting $\mathpzc{O}(m^{4})$ and choosing 
$I=1,\,-1$, i.e.  $m_{1}=1/A$ and $m_{2}=-1/A$, and  
$I=3,\,1$, i.e.  $m_{1}=3/A$ and $m_{2}=1/A$, the logarithm of the two yield ratio can be written as:
\begin{equation}\label{eq:lnRlnR}
\left.
\begin{array}{c}
lnR(1,-1,A)\approx 2\,A\frac{H}{T}\frac{1}{A} \\
\vspace*{-2mm}\\
lnR(3, 1,A)\approx -A\left[\frac{1}{2}a \frac{8}{A^{2}}-2\frac{H}{T}\frac{1}{A}\right] \\
\end{array}\right.  
\end{equation}
where $1/2a$, fitting parameter defined in Eq. \ref{FT}, is by definition $C_{sym}/T$ \cite{bonasera2008,huang2010Theory,huang2010Mscaling,rahul}.
The difference between the two relations, as in Eq. \ref{cymsIYRLandau}, cancels out the dependence on $\frac{H}{T}$, removing the dependence on the characteristics of the source.

We would like to stress here that the main difference between the first two methods of extracting $C_{sym}/T$ and the third is that in the first two cases we are computing  the yield ratios of the same fragment produced by two different sources, while in the third case we consider different isobars produced by a common source. Therefore, in the first two cases the determination of $C_{sym}/T$ requires an estimation of $\frac{\Delta H}{T}$, while in the third case the dependence on the source characteristics is removed by computing the difference of two yield ratios.
However, the isobaric yield ratio method is dependent on the selected isobars and vulnerable to secondary decay effects. Isoscaling and m-scaling are less affected by secondary de-excitation effects, since they are removed, to first order, when taking the ratio of the yields of the same fragment produced in two sources.
We will show that the three methods, which are consistent within the Landau free energy approach \cite{huang2010IYR,chen10}, give  $C_{sym}/T$ in good agreement. Moreover the agreement of the results could suggest the method of determination of $\Delta$.\\

\section{Motivation of the present work}\label{sec:motivation}
In the present work, fragment yield data from quasi-projectile fragmentation in $^{64}$Zn+$^{64}$Zn, $^{70}$Zn+$^{70}$Zn and $^{64}$Ni+$^{64}$Ni  at $35$A MeV have been analysed using the isoscaling, the m-scaling and the isobaric yield ratio methods. There are three issues for the determination of the $C_{sym}/T$ values in Eq. \ref{csym} and \ref{cymsIYRLandau}: the importance of the experimental conditions, the effect of the secondary decay and the method of determination of $\Delta$.

 In a previous work, S.~Wuenschel et al.\cite{wuenschel2009} have presented  data from quasi-projectile fragmentation in $^{78,86}$Kr+$^{58,64}$Ni at $35$A MeV, showing an independence of the isoscaling-extracted $C_{sym}/T$ values from the charge of the analysed fragments. Stringent constraints were applied for a careful characterization of the source with a well defined  neutron (proton) concentration. The data were collected by a $4\pi$ charged-particle detector array.\\
In the works presented in  \cite{huang2010Mscaling,huang2010IYR,chen10}, data for the reactions $^{64,70}$Zn, $^{64}$Ni+$^{58,64}$Ni,$^{112,124}$Sn, $^{197}$Au,  $^{232}$Th at $40\,$AMeV  analysed with isoscaling, m-scaling and isobaric yield ratio methods instead show a strong dependence of $C_{sym}/T$ values on the charge of the analysed fragment. The data were collected with a detector telescope placed at $20^{\circ}$. The telescope allowed the clear identification of typically $6-8$ isotopes for atomic number, $Z$, up to $Z=18$. In contrast to the experimental conditions of S.~Wuenschel et al. \cite{wuenschel2009}, the information on the fragmenting source was not available  and the angular coverage was limited ($15^{\circ}<\theta<25^{\circ}$). Thus the yield of each isotope was evaluated using a moving source fit, selecting fragments produced in the fragmentation of the overlapping region of the two colliding nuclei.

It should be mentioned here that, if compression effects and/or higher order terms in Eq. \ref{cymsIYRLandau} are important, the isoscaling and the isobaric yield ratio methods might produce different results.  Moreover, a different beam energy and different colliding nuclei could access different densities thereby providing complementary information.  Thus investigations under precise experimental constraints for the source properties are necessary, which is one of the goals of this paper.

Furthermore, the observation of an increasing behavior of $C_{sym}/T$ as a function of the fragment mass raises the question of the effects of the secondary decay process on the observables. Indeed, in previous works, excitation energies of the primary fragments  estimated from the associated light charged-particle multiplicities \cite{marie98,hudan03} suggest that effects of secondary evaporation could be important.
Another possibility could be that we are really measuring an enthalpy rather than the internal symmetry energy and different nuclei could be emitted at different volumes.

The third issue is the expression of the quantity $\Delta$ that is used to extract $C_{sym}/T$ from the isoscaling and m-scaling  parameters. Indeed, while $\Delta$ from Eq. \ref{eq:deltaSource} has been widely used after its introduction \cite{tsang2001_2}, the study of the isospin fractionation suggests that the connection between $C_{sym}/T$ and the scaling parameters should be found in the fragment isotopic asymmetry (Eqs. \ref{eq:delta liquido} and \ref{eq:delta mf}). Whether the event composition (Eq. \ref{eq:delta mf})  or the average composition of each fragment (Eq. \ref{eq:delta liquido}) is more relevant is still an open question, which is specifically addressed in this work.

 In this work we investigate the importance of the source reconstruction and  the calculation of $\Delta$ by comparing the $C_{sym}/T$ values obtained with the three different methods.

\section{The experiment}\label{sec:experiment}

\begin{figure}
\centering
\includegraphics[width=1.0\columnwidth]{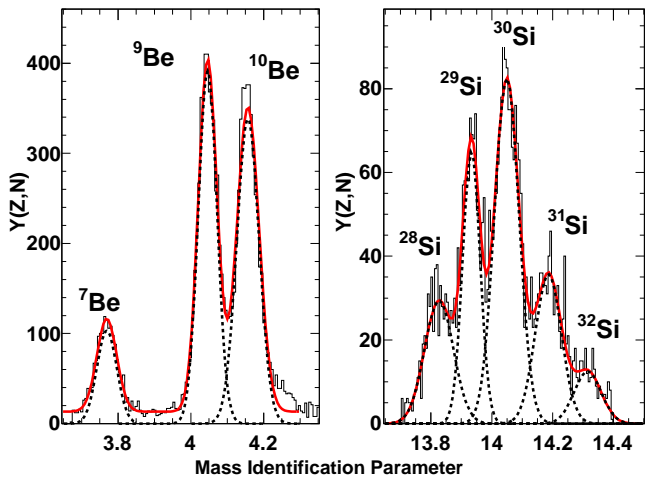} 
\caption{(Color online) Isotopic distributions for $Z=4$ (left) and $Z=14$ (right) for a Super-Telescope. The $x$ axis is obtained by linearizing $\Delta E-E$ spectra as described in \cite{zach}. The gaussian fits (black, dashed) show the overlapping of the mass distributions for different isotopes. The overall gaussian fit is also showed (red, solid) for each $Z$.}
\label{fig:Lx}
\end{figure}

The experiment was performed at Texas A$\&$M University K500 superconducting Cyclotron. Beams of $^{64}$Zn, $^{70}$Zn and $^{64}$Ni at $35\,$AMeV beam energy were  impinged on $^{64}$Zn, $^{70}$Zn and $^{64}$Ni targets. The $4\pi$ NIMROD-ISiS array \cite{wuenschelNimrod, wuenschelNeutronBall} was used for the detection of charged particles, obtaining  isotopic resolution for $Z\leq17$. The detector telescopes, arranged on $15$ rings centered on the beam axis, were composed of one silicon detector backed by  a CsI(Tl) crystal with PMT readout. Two telescopes per ring, referred to as Super-Telescopes, made use of two silicon wafers in front of the CsI(Tl), in order to improve the isotopic resolution. The  charged-particle array was housed inside the TAMU  Neutron Ball \cite{wuenschelNeutronBall}, which measured the free neutron multiplicity.

The mass identification of fragments is essential for isospin physics analysis. For $Z\leq2$, mass identification was performed by pulse shape discrimination of the CsI(Tl) signal. The $\Delta E-E$ method was used for $Z\geq 3$ on Si-Si and Si-CsI telescopes.
In Fig. \ref{fig:Lx} the isotopic distributions of Be and Si are shown as examples.
We can see that the global fit (red, solid)  reproduced the overall spectrum well. The individual gaussian fits (black, dashed) show the overlapping of the mass distributions for different isotopes. 
Particles were confidently assigned a mass if the contamination from neighbour isotopes was less than $5-10\%$. As evidenced by the figure, the mass determination efficiency decreases  for large A near the upper limit of the electronics range. Further details on the mass identification procedures and on  the experiment can be found in \cite{zach}.

Particle and event selections were performed to select quasi-projectile fragmentation events.
The quasi-projectile source was reconstructed accepting in each event fragments with a longitudinal velocity relative to the largest fragment within the range $\pm65\%,\;60\%$ and $45\%$ for $Z=1$, $Z=2$ and $Z\geq3$, respectively \cite{steckmeyer01}. This cut, later referred to as \textit{Vcut}, is intended to remove fragments from non-projectile-like sources. Furthermore the total $Z$ of the detected fragments included in the reconstruction was constrained to be in the range $Z=25-30$ (\textit{SumZ}). Finally limits were placed on the deformation of the source, as measured by the quadrupole momentum, to select a class of events that are, on average, spherical. The quadrupole momentum, calculated from the measured particle momenta in the quasi-projectile frame, $\sum_{i} p_{\|_{i}}^{2}/\sum_{i} p_{\bot_{i}}^{2}$, was required to be less than $2$ (\textit{Qcut}). Details on the source reconstructions can be found in Ref. \cite{wuenschel2010}. Free neutrons measured by the Neutron Ball were used to correct for the free neutrons emitted by the quasi-projectile using the procedure discussed in ref. \cite{wuenschel2010,wuenschel2009}.

\section{Impact of source reconstruction}\label{sec:QP rec}
\begin{figure}
\centering
\includegraphics[width=0.95\columnwidth]{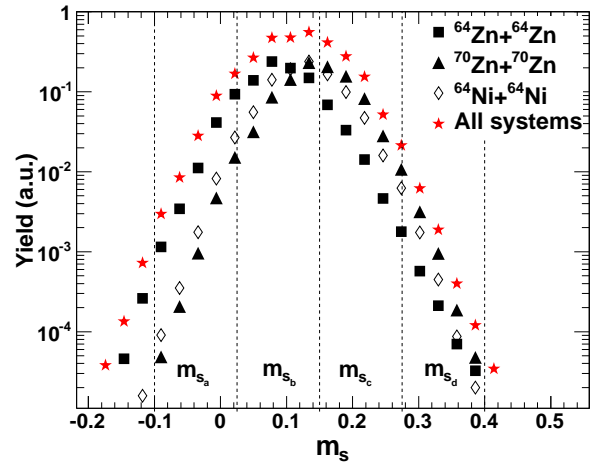} 
\caption{(Color Online) Reconstructed quasi-projectile $m_{s}$ distribution  from the $^{64}$Zn+$^{64}$Zn (squares),$^{64}$Ni+$^{64}$Ni (diamonds), $^{70}$Zn+$^{70}$Zn (triangles) reactions. The total yield of each system was normalized to unity. The $m_{s}$ distribution of the quasi-projectile obtained from the combined systems is plotted as stars. The total yield of this distribution was normalized to $3$.}
\label{fig:msDistribution}
\end{figure}
Figure \ref{fig:msDistribution} shows the distribution of the reconstructed quasi-projectile isospin asymmetry $m_{s}=\frac{N_{s}-Z_{s}}{A_{s}}$ for each of the three systems and for the three systems combined together. In the $m_{s}$ expression, $N_{s}$, $Z_{s}$ and $A_{s}$ are, respectively, the number of neutrons, protons and total nucleons in the reconstructed quasi-projectile. As expected, the systems $^{70}$Zn+$^{70}$Zn and $^{64}$Ni+$^{64}$Ni, which have very similar $N/Z$ values ($1.33$ and $1.29$ respectively), show almost overlapping $m_{s}$ distributions, with average values $\overline{m_{s}}=0.15$ and $0.13$ respectively. The $^{64}$Zn+$^{64}$Zn $m_{s}$ distribution is shifted toward lower $m_{s}$ values ($\overline{m_{s}}=0.09$), since the system is less neutron rich ($N/Z=1.13$). 
The $m_{s}$ distribution of the three combined systems is centred around $\overline{m_{s}}=0.12$.
The widths of the distributions are large compared to the difference in the average $m_{s}$ between the reacting systems. In Ref. \cite{wuenschel2009} it has been shown that an improved isoscaling can be obtained when selecting narrow bins in $N/Z$ of the fragmenting source, rather than performing a system-to-system isoscaling. In analogy, and extending that work, we  performed both isoscaling and m-scaling between two different $m_{s}$ bins. For the isoscaling analysis using $m_{s}$, $\Delta\overline{m_{s}}$ refers to the difference of the mean $m_{s}$ values in the two $m_{s}$ bins. To allow a comparison of the results from the three different methods, for each $m_{s}$ bin combination the isobaric yield ratio was computed for fragments produced by a source with $\langle m_{s}\rangle =\Delta\overline{m_{s}}\pm 0.0625$.

Fragment yield data  was divided according to the relative isospin asymmetry, $m_{s}$, of the reconstructed source in $4$ bins: $m_{s_{a}}=-0.0375\pm0.0625$, $m_{s_{b}}=0.0875\pm0.0625$, $m_{s_{c}}=0.2125\pm0.0625$ and $m_{s_{d}}=0.3375\pm0.0625$, whose boundaries  are also plotted in Fig. \ref{fig:msDistribution}. 
\begin{figure}
\centering
\includegraphics[width=0.95\columnwidth]{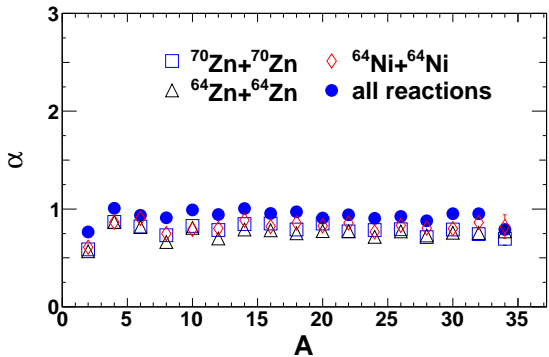} 
\caption{(Color Online) Isoscaling parameter $\alpha$ extracted for $\Delta \overline{m_{s}}=0.185$ and $3.5\leq E^{\star}_{t}/A\leq5\,$AMeV from each reaction and for the combined systems as a function of the fragment mass ($A=2Z$).}
\label{fig:AllReactionsAndEachReaction}
\end{figure}
The transverse excitation energy of the reconstructed source ($E_{t}^{\star}$) was calculated
through calorimetry, as \cite{wuenschel2010,wuenschel2009}:
\begin{equation}
E_{t}^{\star}=\sum_{i}^{M_{CP}}K_{t}^{CP}(i)+M_{n}\langle K_{t}^{n} \rangle -Q.
\end{equation}
The first term is the sum of the transverse kinetic energies in the quasi-projectile center-of-mass of the particles ($M_{CP}$) belonging to the quasi-projectile (see Sec.\ref{sec:experiment} - \textit{VCut}). The free neutrons contribution to the excitation energy was estimated as the average neutron kinetic energy $\langle K_{t}^{n} \rangle$ multiplied by the neutron multiplicity $M_{n}$. The average transverse kinetic energy of the neutrons was calculated as the proton average transverse kinetic energy corrected for the Coulomb barrier energy \cite{dore2000}. The last term in the equation is the reaction $Q$ value. The mass of the quasi-projectile was calculated as the sum of the masses of the charged particles belonging to the source and the neutron multiplicity.
The transverse excitation energy per nucleon ($E_{t}^{\star}/A$) of the reconstructed source was calculated and the data was divided into five bins: $(1.75\pm1.75)$MeV, $(4.25\pm0.75)$MeV, $(5.75\pm0.75)$MeV, $(7.25\pm0.75)$MeV and $E_{t}^{\star}/A>8$MeV.

Figure \ref{fig:AllReactionsAndEachReaction} shows the isoscaling parameter $\alpha$ as a function of the mass fragment $A=2Z$ extracted from each reaction (open symbols) and for the combined systems (full circles). The error bars are comparable to the size of the points. The isoscaling was performed between two sources with different $m_{s}$ ($\overline{m_{s_{1}}}=-0.002$ and $\overline{m_{s_{2}}}=0.183$) so that $\Delta \overline{m_{s}}=0.185$. The source excitation energy was required to be between $3.5$ and $5$AMeV. We observe that consistent $\alpha$ values for each mass were obtained for each system and for all the systems when the data was gated on $m_{s}$. The same results were obtained for all the $m_{s}$ bin combinations and for all the excitation energies. Moreover a similar behavior was observed for $\Delta H/T$ extracted from the m-scaling.
Thus the systems were combined together to increase the statistics.

Significantly better isoscaling and m-scaling were obtained once  two  sources with different $m_{s}$ rather than two different reactions were selected. 
\begin{figure}[b!]
\centering
\includegraphics[width=0.99\columnwidth
]{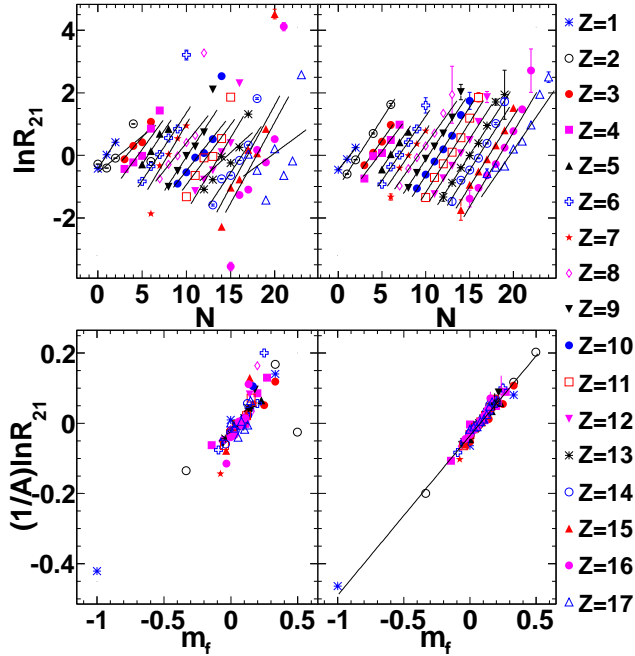} 
\caption{(Color Online) Natural logarithm of the isotopic yield ratio as a function of the fragment $N$ (top) and natural logarithm of the isotopic yield ratio per fragment mass as a function of the fragment isospin asymmetry $m_{f}$ (bottom). Left: system-to-system isoscaling and m-scaling $^{64}$Zn+$^{64}$Zn, $^{70}$Zn+$^{70}$Zn. Right: isoscaling and m-scaling using different $m_{s}$ bins of the reconstructed quasi-projectile (see text).}
\label{fig:MscalingSysToSysAndMs}
\end{figure}
In Fig. \ref{fig:MscalingSysToSysAndMs} the natural logarithm of the yield ratio of Eq. \ref{R12isoscaling} and $\frac{1}{A}ln R_{21}$ of Eq. \ref{R12mirror} are plotted as a function of $N$ and $m_{f}$, respectively. On the left the ratios are computed between the yields of fragments produced in two different reactions ($^{64}$Zn+$^{64}$Zn and $^{70}$Zn+$^{70}$Zn), while on the right the fragments are produced by two sources with  different $m_{s}$ ($\overline{m_{s_{1}}}=0.097$ and $\overline{m_{s_{2}}}=0.180$), to construct the neutron-poor and the neutron-rich sources, as required from Eqs. \ref{R12isoscaling} and \ref{R12mirror}. Comparison of the left and the right panels demonstrates the improvement of isoscaling and m-scaling with a narrowly defined $m_{s}$ source. Indeed the isotopes line up on parallel, equally spaced lines in the isoscaling  and on a single line in the  m-scaling, as predicted by Eqs. \ref{R12isoscaling} and \ref{R12mirror}, respectively.\\

\begin{figure}
\includegraphics[width=1\columnwidth ]
{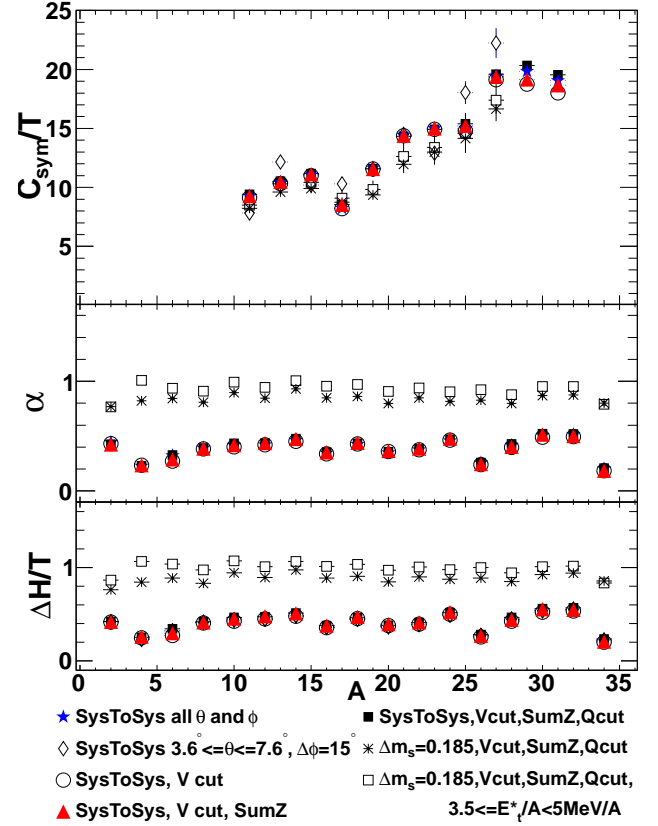}
\caption{(Color Online) $\frac{C_{sym}}{T}$, $\alpha$ and $\frac{\Delta H}{T}$ obtained from isobaric yield ratio (top), isoscaling (middle) and m-scaling (bottom) methods, applied to two different reactions (``SysToSys'') and two $m_{s}$ bins ($\Delta m_{s}$). The source reconstruction constraints are applied to study their impact on the measured values. The error bars, when  not visible, are comparable to the size of the symbols.}
\label{fig:QPeffect_3methods}
\end{figure}

To investigate the impact of the source reconstruction on the extracted values, the source constraints described in this and the previous sections have been imposed sequentially, on all the data, in order to illustrate the effect of each cut individually. The three methods have been applied to the so-selected data.

Figure \ref{fig:QPeffect_3methods} shows $\frac{C_{sym}}{T}$, $\alpha$ and $\Delta H/T$ extracted from the isobaric yield ratio (Eq. \ref{cymsIYRLandau}), the isoscaling (Eq. \ref{R12isoscaling}) and the m-scaling (Eq. \ref{R12mirror}) methods, respectively, as a function of the fragment mass number $A$. For the x-axis of the isoscaling and m-scaling methods $A=2Z$.
The $\alpha$ and $\Delta H/T$ parameters are determined both by individual fits to the yield ratios for isotopes with a given $Z$.\\
We first focus on the values obtained comparing $^{70}$Zn+$^{70}$Zn to $^{64}$Zn+$^{64}$Zn systems (labelled as \textit{SysToSys} in figure), to analyze the impact of a limited angular coverage and of the quasi-projectile reconstruction. We will refer to these values as system-to-system values. A generally flat behavior is observed for  $\alpha$ and $\Delta H/T$, independent of the fragment mass number $A$, when no constraints are imposed on the data (stars). 
The trends are modified neither by restricting our analysis to a limited angular range (diamonds), nor by selecting quasi-projectile fragmentation events (i.e. applying the \textit{Vcut}, \textit{SumZ} and \textit{QCut} -   circles, triangles and full squares, respectively). This suggests that the quasi-projectile selection does not significantly bias our determination of  $\alpha$ or $\Delta H/T$.

In contrast to the flat behavior of the isoscaling and m-scaling, $\frac{C_{sym}}{T}$ increases from $9$ to $20$ over the $A$ range of $11$ to $31$, when no constraints are imposed on the data (stars). The increase is slightly amplified for $A=25$ and $27$ by selecting a limited angular range (diamonds) to simulate the angular coverage of Ref. \cite{huang2010IYR}. The observed increase of $11$ units in $\frac{C_{sym}}{T}$ is similar to the one of $8$ units reported in Ref. \cite{huang2010IYR}. Data for masses $A>27$ in a limited angular range are not available due to a statistics limitation. We should notice that the angular selection does not affect the values extracted from the isoscaling and m-scaling, since we take the ratio of the yields of the same fragment (Eqs.  \ref{R12isoscaling} and \ref{R12mirror}). This is not true for the isobaric yield ratio  method, where different fragments considered in the calculation (see Eq. \ref{R(I+2,I)}) may be affected differently.
To investigate the effect of a good resolution up to $Z=18$, as in Ref. \cite{huang2010IYR}, we restricted the analysis to the detectors with the highest available resolution, the so-called Super Telescopes. Such a restriction does not modify the values of $\frac{C_{sym}}{T}$, therefore it is not plotted in the figure.  Moreover, the selection of quasi-projectile fragmentation events does not modify the $\frac{C_{sym}}{T}$ trend, as can be seen from the figure, since stars (no data selection), circles (\textit{Vcut}), triangles (\textit{Vcut} and \textit{SumZ}) and full squares (\textit{Vcut}, \textit{SumZ} and \textit{Qcut}) overlap.

We turn now to analyse the impact of $m_{s}$ (asterisks) and excitation energy (squares) constraints.
The parameters $\alpha$ and $\Delta H/T$ are plotted for $\Delta \overline{m_{s}} = 0.185$, while $C_{sym}/T$ has been obtained for $\langle m_{s} \rangle=0.183$. The excitation energy was restricted to be between $3.5$ and $5$AMeV. The behavior is flat, for both $\alpha$ and $\Delta H/T$,  and presents less fluctuations with respect to the system-to-system values. The average value is $0.96\pm0.01$, to be compared to the system-to-system average value, $0.34\pm0.01$. The difference in the values observed is due to the different values of $\Delta$ in the two cases. Indeed, the values of $\Delta$, determined as in Eq. \ref{eq:deltaSource} for isoscaling between two different systems and two different $m_{s}$ bins are $0.027$ and $0.084$ respectively, which give consistent values of $\alpha/\Delta$  ($\alpha/\Delta = 12.6\pm0.4$ and $11.43\pm0.12$, respectively). The constancy of $\alpha/\Delta$ was previously shown in \cite{wuenschel2009}.\\
A possible excitation energy dependence of $\alpha$ and $\Delta H/T$ is suggested by systematically higher  (even if in agreement) $\alpha$ and $\Delta H/T$ values obtained for a selected source excitation energy (squares), compared to the ones obtained with no energy selection (asterisks). Indeed the average excitation energy of the source in the latter case is higher ($5.6\pm1.6$AMeV) than the average excitation energy in the selected window ($4.2\pm0.4$AMeV).
This trend is in agreement with the excitation energy dependence of the isoscaling parameter $\alpha/\Delta$ observed for Kr+Ni systems in \cite{wuenschel2009}.

The   selections in $m_{s}$ and excitation energy do not modify the increasing trend of $\frac{C_{sym}}{T}$  extracted by the isobaric yield ratio method. Though the values agree within statistical uncertainty, the $m_{s}$ selection (asterisks and squares) results in a systematic lower $\frac{C_{sym}}{T}$ for $A\geqslant 19$.
\begin{figure}
\centering
\includegraphics[width=0.95\columnwidth]{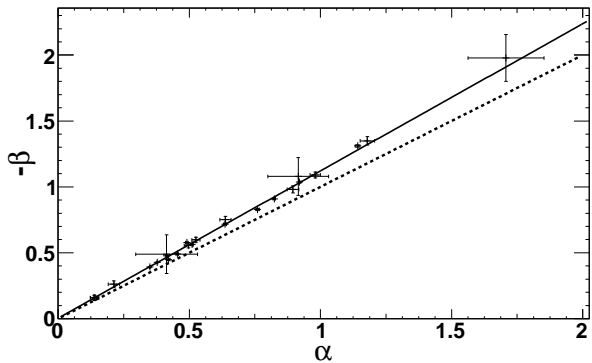} 
\caption{Isoscaling parameters $-\beta$ vs. $\alpha$ for all the  $\Delta\overline{m_{s}}$ and excitation energy bins. The full line is the best fit to the data, while the dashed line represents $\alpha=-\beta$.}
\label{fig:alfavsbeta_fit}
\end{figure}
A possible  excitation energy dependence may be observed also in $\frac{C_{sym}}{T}$  extracted by the isobaric yield ratio method. Indeed for $A\geqslant19$ $\frac{C_{sym}}{T}$ values obtained from data selected in excitation energy (squares) are very slightly higher than those obtained from data not selected in excitation energy (asterisks).
Moreover when performing an excitation energy selection we notice that different mass regions are populated depending on the source excitation energy: higher mass fragments ($A>15$) are produced by low excited sources, while lower mass fragments ($9\leq A\leq15$) are mainly produced by highly excited sources. This is consistent with  fragments  being produced by different reaction mechanisms. 
Selection of heavy fragment multiplicity may weaken the dependence of $\frac{C_{sym}}{T}$  on $A$. However the present statistics do not allow this selection.
These observations suggest that, in the absence of a source reconstruction, attention should be paid to mixing reaction mechanisms  when drawing general conclusions.

\section{Comparison of the three methods}\label{sec: comparison}
The following analysis was performed applying the described selections on the reconstructed quasi-projectile (see Sec. \ref{sec:QP rec} for details). The quasi-projectiles were divided into bins based on their composition $m_{s}$ and their excitation energy per nucleon.

The isoscaling parameters $\alpha$ and $\beta$ were extracted  simultaneously by a global fit to the yield ratios of all the available isotopes and, separately, by individual fits to the yield ratios for isotopes with a given $Z$. The obtained values were in good agreement.
The parameter $\beta$ shows the same trend as a function of $N$ as  $\alpha$ does as a function of $Z$, but has the opposite sign. The parameters, evaluated for all possible $m_{s}$ and excitation energy combinations, are plotted in Fig. \ref{fig:alfavsbeta_fit}. The best fit and the line representing $\alpha=-\beta$ are also plotted in the figure. The relation $\alpha=-\beta$ appears to be approximately satisfied in our data ($-\beta=1.11\,\alpha$), thus supporting the equivalence of the isoscaling and the m-scaling. 
Therefore we can use Eq. \ref{csym} to calculate  $\frac{C_{sym}}{T}$ from m-scaling, using $\alpha=\Delta H/T$.
Small differences between $\alpha$ and $-\beta$ might be attributed to residual Coulomb effects. Indeed, although the ratio of the yields of a single isotope is considered, the two sources, whose $m_{s}$ is fixed, might have  different charges. We remind the reader that the \textit{SumZ} selection constrains the source charge to be $Z=25-30$.

\begin{figure}
\centering
\includegraphics[width=0.95\columnwidth]{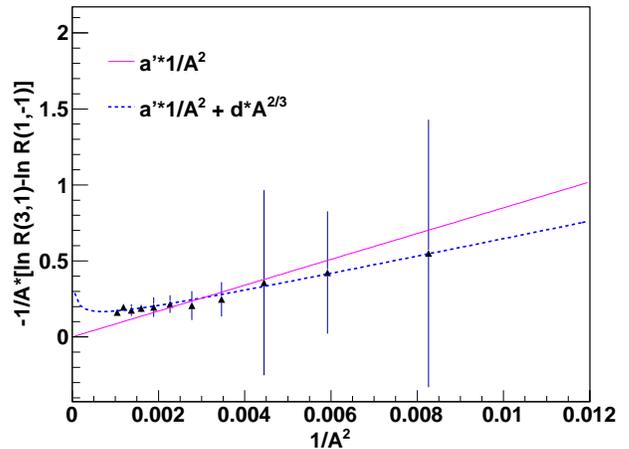} 
\caption{Quantity $-1/A^{2}\left[lnR(3,1)-lnR(1,-1)\right]$ vs $1/A^{2}$ calculated from isobaric yield ratios for an average source isospin asymmetry $\overline{m_{s}}=0$. The dashed and full lines are the best fits to the data with a first degree function, including and not including the Coulomb contribution, respectively.}
\label{fig:fitFT_fit}
\end{figure}
The residual Coulomb effect due to differences in the considered fragments charge is relevant only in the calculation of $\frac{C_{sym}}{T}$ by the isobaric yield ratio method and it influences mainly large mass fragments. 
From the mass formula the Coulomb energy for large $Z$ can be written as \cite{huang2010Theory}:\\
\begin{equation}
\frac{E_{c}}{A} = 0.77\frac{Z^{2}}{A^{2}}A^{2/3} = \frac{0.77}{4}(1-m)^{2}A^{2/3}
\end{equation}
Adding this term to the free energy $\Psi/T$ in Eq. \ref{FT}, we see that a quadratic and a linear term in $m$ are introduced that modify the symmetry energy coefficient and the external field. Also a term not dependent on $m$ is introduced. The effect of introducing such corrections is extensively discussed in Ref. \cite{huang2010Theory}. Here we concentrate on evaluating the Coulomb contribution, as well as the importance of $o(m^{4})$ and $o(m^{6})$ terms in Eq. \ref{FT}. 
First of all we should notice that, assuming a spherical expansion, at low densities the Coulomb energy  decreases as $\rho^{1/3}$. 
A fit of the quantity $-1/A^{2}\left[lnR(3,1)-lnR(1,-1)\right]$ allows us to estimate the fitting parameters of Eq. \ref{FT} and the Coulomb term.
The fit results are shown in Fig. \ref{fig:fitFT_fit}. The dashed and full lines are the best fit to the data with a first degree polynomial of $1/A^{2}$ including and not including the Coulomb correction, respectively. We remind the reader that $1/A^{2}\propto m^{2}$. The parameters $a'$ and $d$ are fitting parameters. We found that $o(m^{4})$ are negligible, since we are dealing with relatively large fragments, which imply small $m$ (but possibly large Coulomb). The Coulomb parameter $d$ is about two orders of magnitude smaller than the $1/A^{2}$ coefficient $a'$, but the term has to be included in the fitting function to obtain a good fit ($\chi^{2}/ndf \approx 0.8$). We therefore took into account this correction and modified Eq. \ref{cymsIYRLandau} as:\\
\begin{equation} \label{cymsIYRLandauCoul}
\left.
\begin{array}{c}
\hspace{-1.5cm}\frac{C_{sym}(T)}{T}\approx -\frac{A}{8}\left[ lnR(3,1,A) -\right.\\
\vspace*{-2mm}\\
\left. \hspace{1.4cm}-lnR(1,-1,A)- \delta(3,1,A)\right]-d\,A^{2/3}
 \end{array}\right.   
\end{equation}

\begin{figure}
\centering
\includegraphics[width=0.95\columnwidth]{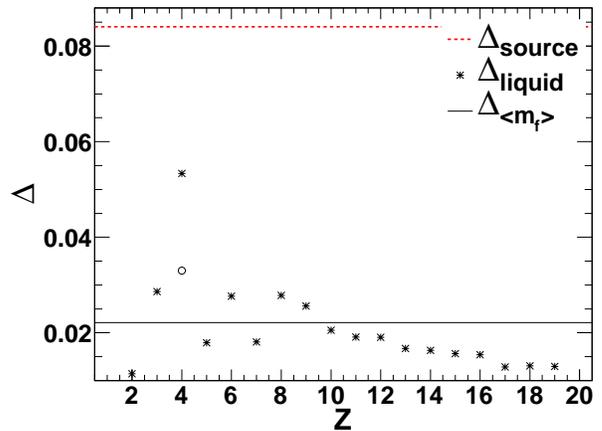} 
\caption{(Color Online) Values of $\Delta$ computed according to the different definitions of Eqs. \ref{eq:deltaSource}, \ref{eq:delta liquido} and \ref{eq:delta mf}, vs. the fragment charge $Z$ for   $\Delta \overline{m_{s}}=0.185$ and $3.5\leq E^{\star}_{t}/A\leq5\,$AMeV.}
\label{fig:Cfr3Delta}
\end{figure}

The $\alpha$ parameters are related to $\frac{C_{sym}}{T}$ through the value of $\Delta$, as given in Eq. \ref{csym}. Therefore, the extracted value  of the symmetry term depends on the choice of $\Delta$. The isobaric yield ratio method instead provides a $\frac{C_{sym}}{T}$ value independent of the choice of $\Delta$. Three different definitions of $\Delta$ have been examined. The $Z/A$ of the emitting source has been adopted for the definition of  $\Delta_{source}$ (Eq. \ref{eq:deltaSource}). The average proton content of each fragment with mass greater than $4$  has been used to calculate  $\Delta_{liquid}$  (Eq. \ref{eq:delta liquido}). Finally the average of the fragment $\langle\overline{m_{f}}\rangle$ over the events has been used to define  $\Delta_{\langle m_{f}\rangle}$ (Eq. \ref{eq:delta mf}).  The three different $\Delta$ values as a function of the fragment charge for a given $\Delta\overline{m_{s}}$ and excitation energy combination  are plotted in Fig. \ref{fig:Cfr3Delta}. 
By definition, $\Delta_{source}$ (red, dashed) and $\Delta_{\langle m_{f}\rangle}$ (black, full) are independent of $Z$. We observe that   $\Delta_{\langle m_{f}\rangle}\approx 0.3 \Delta_{source}$, which is close to the value reported in Ref. \cite{rahul}. 
The large discrepancy between the value of $\Delta_{source}$ and the values of $\Delta_{liquid}$ and $\Delta_{\langle m_{f}\rangle}$ is due to the fact that, while both $\Delta_{liquid}$ and $\Delta_{\langle m_{f}\rangle}$ are related to the fragment composition, $\Delta_{source}$ is related to the source composition.
 The quantity $\Delta_{liquid}$ (asterisks) shows an odd-even behavior for $Z\leq8$, which can be attributed to structure effects \cite{winchester2000}. Such odd-even effect is in agreement with the trend reported in \cite{winchester2000,martin2000,shetty03}. The overall behavior shows a rather significant decrease as a function of $Z$, which will cause  $\frac{C_{sym}}{T}$ to increase as a function of $Z$.  
The very high value of $\Delta_{liquid}$ for Be is due to the fact that $^{8}$Be, produced in the multifragmentation  process, decays before reaching the detector, thus $\langle A \rangle$ deviates from the ``true'' centroid of the isotope distribution. An attempt was made to estimate the $^{8}$Be yield from the $N=Z$ nuclei yields, whose corrected value is plotted in the figure as a circle. A plot of the yield vs. mass number for $m_{f}=0$ fragments displays a power law behavior with $Y\propto A^{-\tau}$ \cite{huang2010Theory}. The yield of $^{8}$Be has been estimated from a fit of $CA^{-\tau}$ to the data for even-even $N=Z$ nuclei, for which the pairing contribution is the same. The  value of $\Delta$ changes from $0.053$ to $0.033$.  
\begin{figure*}[h!]
\centering
\includegraphics[scale=0.75]
{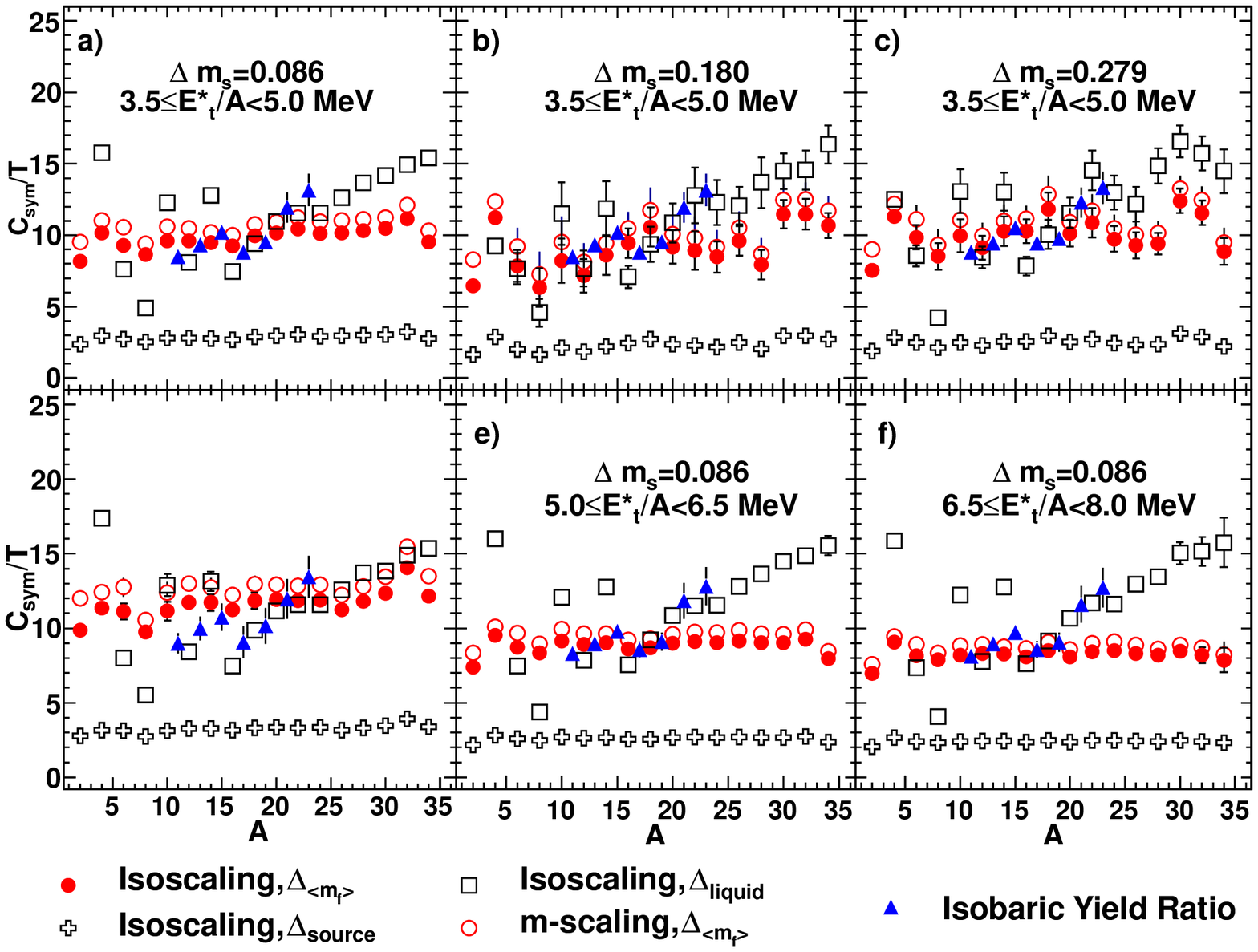}
\caption{(Color Online) Values of $\frac{C_{sym}}{T}$ obtained with the three different methods. Each panel corresponds to the noted $\Delta\overline{m_{s}}$ and the noted window in $E^{\star}_{t}/A$. Three different choices of $\Delta$ have been used to compute $\frac{C_{sym}}{T}$ from the isoscaling parameter $\alpha$.}
\label{fig:CfrMethodsAllDeltaEfixAllBin}
\end{figure*}

Figure \ref{fig:CfrMethodsAllDeltaEfixAllBin} 
 shows a comparison between the three methods, with different choices of $\Delta$, for a given excitation energy ($3.5\leq E^{\star}_{t}/A <5.0\,$AMeV - top panels) and for a given source asymmetry difference ($\Delta\overline{m_{s}}=0.086$ - \textit{a} and \textit{d-f} panels). 
The $\frac{C_{sym}}{T}$ values obtained by the isobaric yield ratio method (triangles) shows an approximately constant behavior for $A<20$, while an increase is observed for $A=21$ and $23$. This is observed for all the $\Delta \overline{m_{s}}$ combinations and excitation energy windows. The observed increase of $3-4$ units in $\frac{C_{sym}}{T}$ for $A$ from $11$ to $23$ is smaller than the increase observed in Ref. \cite{huang2010IYR}. We would like to emphasize that one of the main limitations of the isobaric yield ratio method is that it requires a nearly ``ideal'' isotopic identification, since any contamination by other isotopes to each (Z,A) fragment affects different isobars by a different amount. This effect is particularly relevant for the yield estimate of heavy particles. Indeed the mass resolution decreases with increasing fragment size (see Fig. \ref{fig:Lx}) lowering the particle identification efficiency. While this effect largely cancels out when the yield ratio of the same fragment produced by two different sources is taken (i.e. isoscaling and m-scaling), the $\frac{C_{sym}}{T}$ estimation for the isobaric yield ratio method remains affected. For this reason data for $A\geq25$, when available, are not shown for the isobaric yield ratio method, while they are presented for the isoscaling and m-scaling methods. Indeed, we do not observe a change of behavior of $\alpha$ as a function of A extracted from the isoscaling and m-scaling methods even in the high mass region (see Fig. \ref{fig:AllReactionsAndEachReaction}).

The trends of $\frac{C_{sym}}{T}$ obtained from the isoscaling and m-scaling depend on the definition of $\Delta$, as discussed in Eq. \ref{csym}. First, we notice that $\frac{C_{sym}}{T}$ obtained from the isoscaling with the three definitions of $\Delta$ are not in agreement. This is due to the differences in the values of $\Delta$, which is clearly shown in Fig. \ref{fig:Cfr3Delta}. The $\frac{C_{sym}}{T}$ values determined using $\Delta_{source}$ (crosses) and $\Delta_{\langle m_{f}\rangle}$ (circles) do not show any $A$ dependence, while a clearly increasing trend is observed for $A>15$ for values determined using $\Delta_{liquid}$ (squares). This is caused by the dependence of $\Delta_{liquid}$ on $A$, as shown in Fig. \ref{fig:Cfr3Delta}. For $A<15$ the structure effects affecting $\Delta_{liquid}$ are reflected in $\frac{C_{sym}}{T}$ values.\\
The $\frac{C_{sym}}{T}$ values determined using $\Delta_{\langle m_{f}\rangle}$ from the isoscaling (full circles) and m-scaling (open circles) show a generally good agreement for all the $\Delta \overline{m_{s}}$ combinations. We can see that the values obtained from the m-scaling are systematically higher than those obtained from the isoscaling, which could be due to the residual Coulomb effects noted above. The difference decreases as the excitation energy increases, as can be seen comparing panels \textit{a, d, e} and \textit{f}.

Comparing the top panels, we observe that the values obtained with the isoscaling and m-scaling, using  $\Delta_{\langle m_{f}\rangle}$, do not depend on the choice of $\Delta \overline{m_{s}}$. Indeed, the weighted average values ($10.19\pm0.04$, $10.10\pm0.20$, $10.12\pm0.16$ and $9.22\pm0.05$, $8.90\pm0.20$, $9.07\pm0.15$ for $\Delta \overline{m_{s}}=0.086$, $0.180$ and $0.279$ and extracted by m-scaling and isoscaling, respectively) are consistent within $2\sigma$. The increase in the uncertainties reflects the decrease of available statistics as the considered $m_{s}$ bins are more apart, i.e. when we consider the tails of the $m_{s}$ distribution (see Fig. \ref{fig:msDistribution}).

An excitation energy dependence of $\frac{C_{sym}}{T}$ can be extrapolated from panels \textit{a, d-f}: $\frac{C_{sym}}{T}$ decreases from $11.1 (\pm0.08)$ to $7.7 (\pm0.04)$ as the source excitation energy increases. This trend is observed for the $\Delta_{\langle m_{f}\rangle}$ derived quantities, as well as for $\frac{C_{sym}}{T}$ derived with the other $\Delta$ definitions. As already discussed, this  trend is in agreement with the excitation energy dependence of the isoscaling parameter $\alpha/\Delta$ observed in \cite{wuenschel2009}.
An excitation energy dependence of $\frac{C_{sym}}{T}$ is also observed for the m-scaling extracted quantities, which vary from $12.47(\pm0.08)$ to $8.20(\pm0.04)$ increasing the excitation energy. In contrast, a weaker excitation energy dependence is found for the values extracted with the isobaric yield ratio method, which decreases from $10.0(\pm0.4)$ to $8.8(\pm0.2)$ increasing the excitation energy.

We now compare the values obtained with the different $\Delta$ definitions to the values extracted from the isobaric yield ratio, since they are independent on the choice of $\Delta$.
The $\frac{C_{sym}}{T}$ values determined using $\Delta_{source}$ (crosses) are lower than those determined with the isobaric yield ratio method (triangles) by a factor $2$ to $3$ independent of $\Delta m_{s}$ and independent of the excitation energy. The $\frac{C_{sym}}{T}$ values determined using $\Delta_{liquid}$ (squares) show a rather good agreement for $A\approx18-22$ in the central mass region, where the $\Delta_{liquid}\approx \Delta_{\langle m_{f}\rangle}$. Nevertheless, it does not reproduce the data for the low mass region, where structure effects influence the value of $\Delta_{liquid}$, as pointed out in Fig. \ref{fig:Cfr3Delta}. In this region, in contrast to the high mass region, the isotopic resolution of each fragment is very good and presents a small uncertainty; thus, we are confident on the extracted $\frac{C_{sym}}{T}$ values. The $\frac{C_{sym}}{T}$ values determined using $\Delta_{\langle m_{f}\rangle}$, both from isoscaling and m-scaling, show  good agreement with the isobaric yield ratio values for masses up to $A\approx20$, for all the $\Delta \overline{m_{s}}$ and excitation energy combinations. This is consistent with what was recently observed by  R. Tripathi et al. in Ref. \cite{rahul}. In this work the authors show that the position of the free energy central minimum, when an external field is present (i.e. when $m_{s}\neq0$ \cite{huang2010Mscaling}), is related to  $\langle \overline{m_{f}}\rangle$ rather than to $m_{s}$. This indicates that the connection between quantities extracted from fragment yields and the symmetry energy coefficient has to be found in the composition of the emitted fragments, taking into account  the event multiplicity.

The generally flat behavior of  $\frac{C_{sym}}{T}$ as a function of $A$ (circles and triangles in Fig. \ref{fig:CfrMethodsAllDeltaEfixAllBin}) may be interpreted as the weak influence of  secondary decay effects on the observables. This in turn implies that fragments are determined in the instability region in agreement with the theoretical work of Dorso and Randrup \cite{dorso93}. This would also be the case in a first-order phase transition for infinite nuclei, where the size of a cluster is determined by its internal pressure which takes into account the surface tension, and the external pressure due to the gas \cite{huang_book}. For finite systems  the formed fragments might reach the ground state by emitting low-energy gamma rays and possibly a neutron.  
Also the constancy of $\frac{C_{sym}}{T}$ might indicate that the assumption of constant volume is reasonable and we can identify this quantity as the symmetry energy rather than the enthalpy.

\section{Conclusions}\label{sec:conclusions}
Three methods to extract the symmetry energy coefficient from fragment yields were compared. The isobaric yield ratio method removes the dependence on the fragmenting source by computing the difference of the yield ratios of properly chosen isobars produced by the same source, but the dependence on individual isobar detection efficiencies remains. The m-scaling and the isoscaling, however, retain a dependence on the source characteristics in the difference of the external fields of the two sources ($\Delta H/T$), while significantly reducing the dependence on isotopic detection efficiencies. The determination of $\frac{C_{sym}}{T}$ in this case depends on the choice of $\Delta$.

The symmetry energy coefficient to temperature ratio, $\frac{C_{sym}}{T}$, was experimentally evaluated as a function of the fragment mass with  three different methods. The effect of the source reconstruction on the $\frac{C_{sym}}{T}$ values was analysed, showing that   $4\pi$ angular coverage is useful in extracting information from the isobaric yield ratios, since a limited coverage impacts $\frac{C_{sym}}{T}$ values. 
Improved isoscaling and m-scaling were observed when selecting  quasi-projectile $m_{s}$ bins for n-rich and n-poor systems, while isobaric-yield-ratio-extracted values were only slightly affected. A decrease in $\frac{C_{sym}}{T}$ was observed with increasing excitation  energy.

The isoscaling parameters $\alpha$ and $\beta$ were extracted as a function of $Z$ for all $m_{s}$ and excitation energy combinations. Our data show the equivalence of the isoscaling and the m-scaling, since the relation $\alpha=-\beta$ is approximately satisfied.

The $\frac{C_{sym}}{T}$ values extracted from the $\alpha$ isoscaling parameter using $\Delta_{\langle m_{f}\rangle}$ are in good agreement with the isobaric yield ratio $\frac{C_{sym}}{T}$ values. It indicates that the connection between $\alpha$ and $\frac{C_{sym}}{T}$ has to be found in the average fragment isotopic asymmetry.
The extracted symmetry term shows a generally flat trend as a function of $A$ for the mass region $A=10-20$ with the best mass resolution, independent of the method used. This may be consistent with the lack of secondary de-excitation effects in our data, within our isotopic resolution, as well as with the assumption of a freeze-out volume where the disassembly occurs.

\section*{Acknowledgements}
This work was supported by the U.S. DOE grant DE-FG03-93ER40773 and the
Robert A. Welch Foundation grant A-1266.
\bibliography{../bibliogr}
\bibliographystyle{unsrt}

\end{document}